\documentclass[aps,prl,preprint,groupedaddress]{revtex4}

\usepackage{graphicx}

\def\gamnas{{Ga$_{1-x}$Mn$_{x}$As}}
\def\gacras{{Ga$_{1-x}$Cr$_{x}$As}}
\def\gacrn{{Ga$_{1-x}$Cr$_{x}$N}}

\begin{document}


\title{Role of Disorder in Mn:GaAs, Cr:GaAs, and Cr:GaN}



\author{J.~L. Xu}
\affiliation{Arizona State University, Tempe, AZ, 85284}

\author{M. van Schilfgaarde}
\affiliation{Arizona State University, Tempe, AZ, 85284}
\email[]{Mark.vanSchilfgaarde@asu.edu}

\author{G.~D. Samolyuk}
\affiliation{Ames Laboratory, Iowa State University, Ames, IA, 50011}

\date{\today}

\begin{abstract}

We present calculations of magnetic exchange interactions and
critical temperature $T_c$ in \gamnas, \gacras\ and \gacrn.  The
local spin density approximation is combined with a
linear-response technique to map the magnetic energy onto a
Heisenberg hamiltonion, but no significant further approximations
are made.  Special quasi-random structures in large unit cells
are used to accurately model the disorder.  $T_c$ is computed
using both a spin-dynamics approach and the cluster variation
method developed for the classical Heisenberg model.

We show the following: $(i)$ configurational disorder results in
large dispersions in the pairwise exchange interactions; $(ii)$
the disorder strongly reduces $T_c$; $(iii)$ clustering in the
magnetic atoms, whose tendency is predicted from total-energy
considerations, further reduces $T_c$.  Additionally the exchange
interactions $J(R)$ are found to decay exponentially with
distance $R^3$ on average; and the mean-field approximation is
found to be a very poor predictor of $T_c$, particularly when
$J(R)$ decays rapidly.  Finally the effect of spin-orbit coupling
on $T_c$ is considered.  With all these factors taken into
account, $T_c$ is reasonably predicted by the local spin-density
approximation in MnGaAs without the need to invoke compensation
by donor impurities.

\end{abstract}


\maketitle

\def\tc{$T_c$}

Dilute magnetic semiconductors (DMS), i.e. semiconductors doped
with low concentrations of magnetic impurities (usually Cr, Mn,
or Co), have attracted much interest because of their potential
application to spintronics\cite{Ohno96,Ohno99} .  \gamnas\ is the most
widely studied DMS, and it continues to attract interest because
it is one of the few DMS where it is generally agreed that the
magnetism is carrier-mediated.  (This is important in spintronics
because the magnetic state can be manipulated by electrical or
optical means.)

In recent years Curie temperatures in \gamnas\ have risen
steadily, reaching $\sim$170K for $x${}$\sim$0.08 when grown in
thin films annealed at low temperature
\cite{Chiba03,Ku03,Edmonds04}.  It is generally believed defects
(probably Mn interstitials) migrate out of the as-deposited films
during the anneal, largely eliminating donor defects that hamper
ferromagnetism.  Since most practical applications of spintronics
require room-temperature operation, a crucial question is then,
what is the ultimate limit to $T_c$ in the DMS compounds, and in
\gamnas\ in particular?

This question was first addressed by Dietl in his now classic
paper\cite{Dietl00}, where he predicted a wide range of $T_c$ in
tetrahedrally coordinated alloys.  This stimulated a great deal
of interest, although there is a growing consensus that most of
the claims of that paper were artifacts of the assumptions in his
original model.  On the other hand, Akai\cite{Akai98} first used
the local spin-density approximation (LSDA) to estimate $T_c$
within the Coherent Potential Approximation (CPA) in (In,Mn)As;
he argued that a double exchange mechanism was a more appropriate
description of the magnetism than the $pd$ exchange assumed by
Dietl.  Since then LSDA calculations of exchange interactions
have been performed by a variety of groups
\cite{mark01,Sandratskii02,Bouzerar03,Kudrnovsky04,Erwin03,Sato03},
usually extracting exchange parameters by calculating total
energies of a fixed atomic but multiple-spin configurations, or
by a linear-response technique within the CPA.

To date, disorder has almost always been neglected or treated
within some mean-field (MF) approximation (MFA), either in the
computation of the exchange parameters themselves, or in the
subsequent analysis of magnetization $M(T)$ at
finite-temperature, or both (though better treatments 
within $k\cdot p$ theory has been
reported \cite{Schliemann01}).  The LSDA+MF predict a rather high
$T_c$ for \gamnas\ (typically 350$\sim$400~K for $x${}$\sim$0.08 \cite{Sato03}).
The large discrepancy with experiment (at least in Mn:GaAs) is
usually attributed to the very large numbers of compensating
defects in real samples, which reduce $T_c$
\cite{Chiba03,Ku03,Edmonds04}.  The situation remains somewhat
uncertain because the number of defects still remaining in the
best samples to date is not known.

This Letter addresses the issue of the ultimate limit to $T_c$ in
some DMS alloys (focusing on Mn:GaAs) by adopting relatively
rigorous approach to the calculation of the magnetic exchange
interactions and $T_c$.  Random alloys are approximated by large
(128-250 atom) supercells where special quasirandom structures
(SQS) \cite{zunger90} are used for the cation sublattice.  Using
a linear-response technique within the LSDA and the
linear-muffin-tin orbitals method\cite{licht87,mark99}, the
magnetic energy is mapped onto a Heisenberg form\cite{notea}
\begin{equation}
  H = -\sum_{ij} J(R_{ij}) \, {\hat e_i}\cdot{\hat e_j}
\label{eq:heisenberg}
\end{equation}
where the sum is over all pairs $ij$ of magnetic atoms.  To model
$M(T)$ and $T_c$, Eq.(\ref{eq:heisenberg}) is treated classically
and integrated using a spin-dynamics (SD) technique\cite{sdyn96};
alternatively $M(T)$ is estimated by the cluster variation method
(CVM)\cite{kikuchi} adapted\cite{notecvm} to solve
Eq.(\ref{eq:heisenberg}).  Thus it is evaluated without recourse
to empirical parameters or to the MFA.  We show that the widely
used MFA turns out be a very poor predictor of $M(T)$ in these
disorded, dilute alloys, dramatically overestimating $T_c$.

With SQS we can rather precisely mimic a fully random
configuration, but it is also possible to consider configurations
that deviate from random.  This can be important because LSDA
predicts a strong attractive interaction between magnetic
elements \cite{mark01}, which implies a tendency towards
clustering.  In brief, we show that
\begin{itemize}
\item
      the disorder induces large fluctuations in
      $J_{ij}\equiv{}J(R_{ij})$ for every connecting vector
      $R_{ij}$;
\item The fluctuations in $J_{ij}$ {\em{reduce}} $T_c$ relative
      to the configurationally averaged $\overline
      J_{ij}=\left<J_{ij}\right>$;
\item clustering {\em{reduces}} $T_c$, while ordering of the
      magnetic elements {\em{increases}} $T_c$.
\end{itemize}

\begin{figure}[ht]
\centering
\includegraphics[width=8.5cm]{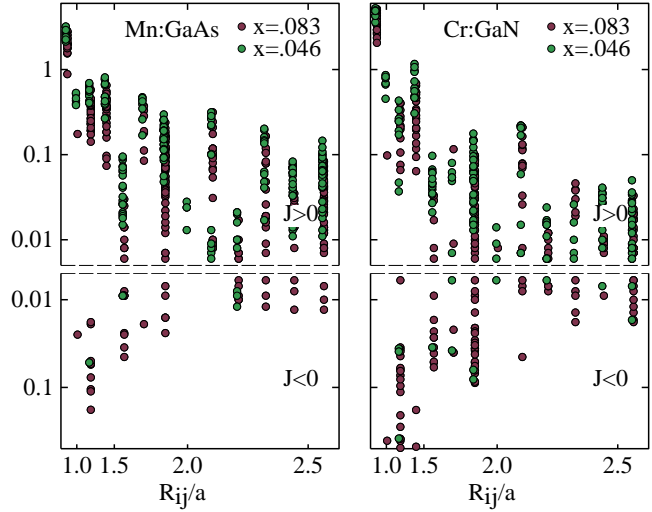}
\caption{Pair exchange interactions $J(R_{ij})$, in mRy, for Mn:GaAs and
  Cr:GaN at two different concentrations as a function of
  $R_{ij}^3$. $R_{ij}$ is measured in units of the lattice
  constant $a$.}
\label{fig:jij}
\end{figure}

Fig.~\ref{fig:jij} shows $J_{ij}$ computed for an ensemble of
108-cation (216-atom) random supercells following the method of
Ref.\cite{mark99}, for \gamnas\ and \gacrn\ alloys at $x$=4.6\%
and $x$=8.3\%.  $3\times3\times3$ $k$-points were used, enabling
the calculation of $J$ to very distant neighbors.  We chose these
two alloys because they are approximately representative of
limiting cases.  For Cr:GaN, the GaN host has a wide bandgap, and
the Cr $t_2$ level falls near midgap.  It broadens into an
impurity band with 1/3 occupancy, and is believed to be
responsible for the ferromagnetic exchange.  For Mn:GaAs, most of
the weight of the Mn $t_2$-derived state falls below the valence
band maximum.  A second $t_2$ impurity band about 0.1~eV above
the valence band maximum is mainly responsible for the
ferromagnetic exchange coupling in this case; the strength of
$J(R)$ depends critically on the amount of Mn character in this
band\cite{Mahadevan04}.  Katayama-Yoshida used the $x$-dependence
of $J_0=\sum_R\overline J(R)$ (computed within the CPA) to
identify the ferromagnetism obtained from LSDA with model
theories\cite{Sato03}.  Within the CPA, $J_0\sim{}x^{1/2}$ for Cr:GaN, which
corresponds to a double-exchange model, while Mn:GaAs displays
character intermediate between $J_0\sim{}x^{1/2}$ and the $pd$
exchange ($J_0\sim{}x$) usually assumed by $k\cdot p$ models
\cite{MacDonald99,Dietl00}.

Comparing Cr:GaN to Mn:GaAs, Cr:GaN shows substantially stronger
nearest-neighbor (NN) interactions, owing to its small lattice
constant; however $\overline J(R_{ij})$ decays much more rapidly
with $R_{ij}$.  This is because the wave function overlap between
transition metal $d$ states decays much more rapidly for midgap states than
near band-edge states.  Evident also is the large dispersion in
$J_{ij}$ for fixed $R_{ij}$ (note $J$ is drawn on a log scale):
the root-mean square fluctuations $\Delta J_{ij} =
\sqrt{\left<J_{ij}^2-\bar{J}_{ij}^2\right>}$ are roughly
comparable to ${\overline J}$.  However $\Delta J_{ij}$ {\em
increases} with $x$, and is substantially larger for the wide-gap
case (Cr:GaN).  Note that there is little evidence in either
Cr:GaN or Mn:GaAs for oscillatory RKKY-like behavior, which in
the simplest approximation predicts $J(R)\sim \cos(2k_F{}R)/R^3$.
Instead, ${\overline{J}(R)}$ decays roughly exponentially in
$R^3$, corresponding to a Fermi surface with imaginary wave
number, as would obtain if the coupling were described by
tunneling via a disordered impurity band\cite{Berciu01}.



We now apply Eq.(\ref{eq:heisenberg}) to compute $M(T)$, focusing
on $T_c$.  Mean-field theory, which estimates the effective field
at each site from the average field contributed by other sites,
predicts $T_c$ well above room temperature both in Mn:GaAs and
Cr:GaN\cite{Newman03,Sato03}.  In spite of the
rather strong differences in the form of $J(R)$
(Fig.~\ref{fig:jij}), mean-field theory predicts that Mn:GaAs and
Cr:GaN have roughly similar $T_c$ for
$x${}$\sim$0.08\cite{Newman03}.  This is because the NN
interaction in the latter case is strongest, but the $J$ decays
faster with $R$, leading to a comparable mean-field\cite{notemf}
estimate $\overline{T}^{MFA}_c$.

But it should be evident from Fig.~\ref{fig:jij} that the MFA is
of questionable reliability.  First, it is well known that for
dilute alloys there is a percolation threshold for the onset of
ferromagnetism.  (The threshold in the present case cannot be
readily mapped to known models because $J(R)$ is nonneglible for
a rather large number of neighbors.)  Moreover, the large
fluctuations $\Delta J(R)$ may strongly affect $T_c$, especially
since $\Delta J(R)$ itself is purely a function of the
environment\cite{mark01}, and consequently of the local
percolation path.

To obtain a precise estimate for $M(T)$ and $T_c$, we adopt a
spin-dynamics approach\cite{sdyn96}.  A 200 atom SQS structure
(250 atom for the 4\% alloy) was used to mimic the random alloy.
From the TM atoms in the SQS structure, a supercell containing
$\sim 2000$ Mn or Cr atoms was constructed to make a simulation
cell for prosecuting spin-dynamical simulations.  Following the
method described in Ref.\cite{sdyn96}, the Landau-Lifshitz (L-L)
equation was integrated numerically at a fixed temperature
allowing the system to equilibrate, followed by a simulation for
$\sim 2\times10^6$ atomic units.  The L-L equations were
integrated with the Bulirsch-Stoer method.  As the L-L equation
is a first-order equation, global deamons were used for the heat
bath\cite{sdyn96}, to ensure ergodic behavior.  The average
magnetization $\overline{M}(T)$ was computed as a function of
temperature, and $T_c$ was estimated from the inflection point in
$\overline{M}(T)$.  Owing to finite-size effects and the
stochastic character of the simulation, $T_c$ could be determined
to a precision of $\sim$5\%.

\begin{figure}[ht]
\centering
\includegraphics[width=6.0cm]{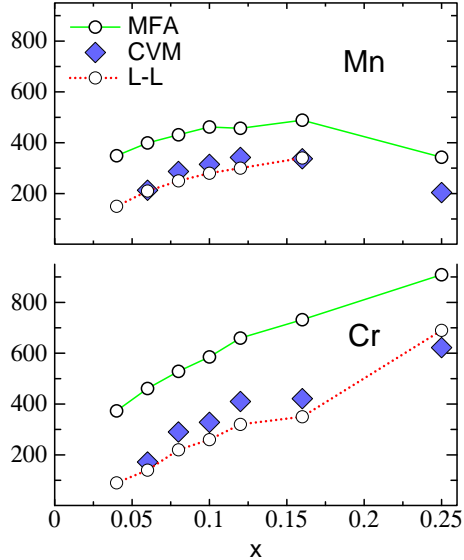}
\caption{Dependence of $T_c$ (K) on $x$ in \gamnas\ and \gacras.
Solid lines: $T_c$ computed from the MF
$\overline{T}^{MFA}_c$\protect\cite{notemf}.  Dotted line:
$T_{c}$ extracted from spin-dynamics simulations of
Eq.~\ref{eq:heisenberg}.  Diamonds: $T_{c}$ computed from the
Heisenberg Cluster Variation Method.
\label{fig:tc}
}
\end{figure}

Also we employed a CVM approach recently adapted to the classical 
Heisenberg hamiltonion\cite{notecvm}.  This
relatively simple scheme has been found to be accurate in simple
3$d$ magnets, overestimating $T_c$ by $\sim$5\% (similar to the
usual CVM for the Ising hamiltonion\cite{vaks99}).  We can check the validity
both methods in the DMS case by comparing their predictions of
$T_c$.  Fig~\ref{fig:tc} shows $T_c$ determined by both methods
for \gamnas\ and \gacras\ : agreement between the two methods is
$\sim$10\%, which is quite satisfactory considering the
complexity of the $J_{ij}$.  $\overline{T}^{MFA}_c$ is also
shown: evidently the MFA rather badly overestimates $T_c$.
$\overline{T}^{MFA}_c>T_c$ by $\sim$200K in the Mn:GaAs alloy,
and by a somewhat larger amount in Cr:GaAs.  The discrepancy is
still more dramatic in Cr:GaN (not shown); we find $T_c<50$K for
all concentrations studied while $\overline{T}^{MFA}_c\sim
600$K\cite{Newman03}.  Indeed we have found this generally to be
the case when $J(R)$ decays rapidly or when $\Delta J(R)/J(R)$ is
not small.

These results stand in stark contrast to the $\sim$15\%
discrepancy between $T^{MFA}_c$ and $T_c$ typically found in
simple metals.  The reason is easily understood by considering
the effective field a mean-field atom sees, $\vec
H^{eff}_i=\sum_jJ_{ij}{\hat e}_j$.  From the exponential decay of
$J(R)$, it is evident that $H_i$ will be dominated by the nearest
neighbors.  But for dilute alloys, near-neighbors are not
sufficent to form a percolation path.  This is immediately
evident in the extreme case of a NN pair of magnetic atoms well
separated from any other magnetic atoms: the contribution to
$T^{MFA}_c$ from this pair would be high, even though the pair
would actually contribute nothing to ferromagnetism.


In Ref.~\cite{Bouzerar03} a small discrepancy between $T^{MFA}_c$
and a more sophisticated calculation for $T_c$ was reported.  In
that calculation the CPA was used to construct an average
$\overline{J}_{ij}$ and $M(T)$ modeled by constructing a fcc
lattice of magnetic atoms, using concentrated-weighted
$\overline{J}_{ij}$ for the exchange parameters.  It would seem
that their conclusions are an artifact of the neglect of
configurational disorder (except in the computation of
$\overline{J}_{ij}$).  Better would be to estimate
$\overline{J}_{ij}$ within the CPA, and then construct a {\em
disordered} simulation cell using the $\overline{J}_{ij}$ to
estimate $M(T)$.  Still this approach neglects fluctuations
$\Delta J$, which as we have seen are comparable to
$\overline{J}_{ij}$ itself.  To assess the effect of
fluctuations, we repeated the calculation for $T_c$ within the
CVM, replacing the environment-specific ${J}_{ij}$ with the
configurationally averaged $\overline{J}_{ij}$.  For \gamnas\ at
$x$=0.08, the effect of disorder ($\overline{J}_{ij}\to{J}_{ij}$)
was to reduce $T_c$ by 50~K.  (It is interesting that the MFA
predicts the {\em opposite} trend, because of an artificial
tendency for $M^{MFA}(T)$ to track whichever site $i$ has the
largest $\vec H^{eff}_i$.  Then $T^{MFA}_c-\overline{T}^{MFA}_c$
is positive\cite{notemf} and increases with $\Delta J/J$.  This
explains why a tight-binding+MF analysis\cite{Berciu01} predicted
that disorder {\em increases} $T_c$.)

We next consider the effects of nonrandomness.  As noted above,
real DMS alloys should exhibit some clustering owing to the
attractive interaction between magnetic elements\cite{mark01}.
The true situation is complicated by the nonequilibrium growth
required to stabilize the alloy in the zincblende structure.
Nevertheless the Mn-Mn or Cr-Cr binding energy is
calculated\cite{mark01} to be an order of magnitude larger than
the growth temperature ($\sim$250K), and some pairing or other
clustering should be expected, particularly since films must be
annealed to obtain good $T_c$.  There is some experimental
evidence for a tendency to cluster\cite{Sullivan03}.

\begin{figure}[ht]
\centering
\includegraphics[width=6.5cm]{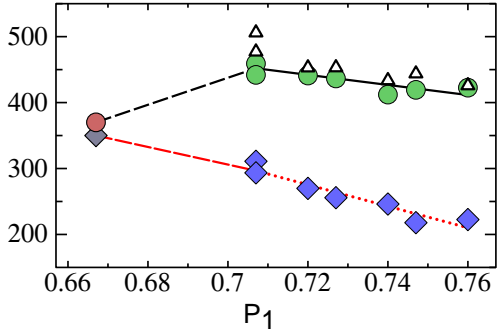}
\caption{Dependence of $T_c$~(K) on the pair correlation function
  $P_1$ in Ga$_{0.92}$Mn$_{0.08}$As.  The random (SQS) configuration
  corresponds to $P_1$=0.7056.  (Two SQS structures were calculated.)
  Diamonds show $T_c$ computed with CVM; circles show
  $\overline{T}^{MFA}_c$, and triangles show ${T}^{MFA}_c$.  The
  point at $P_1=2/3$ corresponds to the ordered compound.
}
\label{fig:tccluster}
\end{figure}

The effect of clustering on $T_c$ in Ga$_{0.92}$Mn$_{0.08}$As was
studied by a simple model.  To characterize the configurational
disorder we adopt the standard Ising formalism, and assign
$\sigma=\pm 1$ to each cation site (+1 for Mn and $-$1 for Ga).
The random (SQS) configuration was constructed by searching for
configurations which best approximate the ideal random
configuration for pair correlation functions
$P_{R_{ij}}=\left<\sigma_i\sigma_j\right>$ (and some higher-order
correlation functions) up to some fixed distance.  For a random
configuration, $P_R=(2x-1)^2$ independent of $R$.  To parameterize
the clustering in a simple manner, we adopted the NN pair 
correlation function $P_1$ as a measure of clustering.  Starting from
an initial SQS configuration, a simulated annealing cycle was
performed by generating a set of site configurations with increasing
$P_1$, corresponding to longer annealing times (For simplicity, 
$P_n(n>1)$ was optimized to be $(2x-1)^2$ for each configuration.)
$J_{ij}$ and $T_c$ were computed by the CVM and MFA\cite{notemf} as a function
of $P_1$; see Fig.~\ref{fig:tccluster}. $T_c$ is rather strongly {\em
reduced} with increasing $P_1$.  This is perhaps not surprising
since increased clustering implies more distant average
separation between atoms, which is deleterious to links in the
percolation path.  Even within the MFA $T_c$ changes slightly,
albeit for a different reason.  In that case, there is an
increase in NN pairs, which would increase $T_c$, but at
the same time there is some increase in the likelihood of {\em
three-} and higher body neighbors.  The presence of a third
neighbor has the effect of {\em reducing} the pairwise $J_{ij}$
\cite{mark01}, and is the origin of the factor-of-three
variations in the NN $J$ in Fig.~\ref{fig:jij}.

We also considered the {\em ordered} limit, by putting 1 Mn in a
24-atom unit cell, corresponding to $x$=0.083.  In this case
$P_1$ decreases to 2/3, and $T_c$ increases to 350K (see
Fig.~\ref{fig:tccluster}). Thus we conclude that ordering {\em
increases} $T_c$, while clustering {\em decreases} $T_c$.
Perhaps not suprisingly, the MFA $T_c$ approaches the CVM result
in the ordered case, since percolation is less critical.

To conclude, we have shown that ferromagnetism is very sensitive
to configurational disorder in DMS alloys, and that with proper
treatment of disorder $T_c$ is reasonably predicted by the LSDA
for \gamnas, without needing to invoke compensating defects.  We
briefly consider two important sources of error from elements
missing in the theory.  First, spin-orbit coupling strongly
reduces $T_c$ in $k\cdot{}p$ models.  We estimated its effect by
computing the change in $\overline{T}^{MFA}_c$ when the
$L\cdot{}S$ coupling is added to the LSDA hamiltonion.  For
Ga$_{0.92}$Mn$_{0.08}$As, $\overline{T}^{MFA}_c$ was reduced by
$\sim$10\%.  Finally, the LSDA itself will overestimate $T_c$
somewhat \cite{Mahadevan04}.  In a future work we will present a
reliable parameter-free theory that corrects the principal errors
in LSDA---most importantly the Mn $d$ character at $E_F$---and
quantify the extent to which the LSDA overestimates $T_c$.
Finally, we conclude that the present calculations represent a
rather strict upper bound to $T_c$, and that for random or
clustered \gamnas\ alloys, $T_c>250$~K is unlikely.


This work was supported by the Office of Naval Research.

\begin{thebibliography}{28}
\bibitem[{not({\natexlab{a}})}]{notea}
\bibinfo{note}{Because the TM magnetic moments are large, the local-moment
  picture and long-wave approximation is expected to be rather accurate; see
  \cite{vladimir03}}.
\bibitem[{not({\natexlab{b}})}]{notecvm}
\bibinfo{note}{In contrast to the Ising model\cite{vaks99}, the classical
  Heisenberg model has continuous rotational degrees of freedom which leads to
  a more complicated form for the partition function: $Z_{ij} =
  \sum_{l}(2l+1)i_l(\psi^{ij}_i/T)i_l(J_{ij}/T)i_l(\psi^{ij}_j/T)$, where
  $i_l(z)=\sqrt{(\pi/2z)}I_{l+1/2}(z)$ and $I_n(z)$ is the modified Bessel
  function of the first kind. Cluster methods can include strong correlation
  among atoms, which the MFA neglects. The CVM accurately describes
  interactions within the clusters (pair clusters in our case). The
  interactions of the cluster with the environment is described by
  renormalizing fields $\psi^{ij}$.}
\bibitem[{not({\natexlab{c}})}]{notemf}
\bibinfo{note}{Mean-field theory for the Heisenberg model approximates the
  coupling to neighboring sites $j$ by an effective field
  $\vec{H}^{eff}_i=\sum_j J_{ij}\hat{e}_j$ at site $i$, and entails the
  solution of a set of coupled nonlinear equations in $H_i$ and $\hat{e}_i$.
  Alternatively, we can replace $J_{ij}$ with its average $\overline{J}_{ij}$
  and then calculate $M(T)$ for the $\overline{J}_{ij}$. We call $T_c$ computed
  by the two approaches respectively, $T^{MFA}_c$ and $\overline{T}^{MFA}_c$.
  The latter can be easily evaluated as $\overline{T}^{MFA}_c=2/3
  \left<\sum_{ij}J_{ij}\right>$.}
\end{thebibliography}

\end{document}